\documentstyle[twoside,fleqn,espcrc2,texdraw]{article}
\title{Lattice Determination of the $B^\ast B \pi$ Coupling
\thanks{Talk presented by Massimo Di Pierro}}
\author{G. M. de Divitiis$^{\mathrm b}$, L. Del Debbio$^{\mathrm a}$, M. Di Pierro$^{\mathrm a}$, J. Flynn$^{\mathrm a}$ and C. Michael$^{\mathrm c}$ [UKQCD Collaboration]
\vskip 5mm
$^{\mathrm a}$ Dept. of Physics and Astronomy, Univ. of Southampton,
Southampton SO17 1BJ, UK \\
$^{\mathrm b}$ Dip. di Fisica, Univ. di Roma ``Tor Vergata'', Via della ricerca scientifica 1, 00133 Roma, Italy \\
$^{\mathrm c}$ Dept. of Mathematical Sciences, Univ. of Liverpool, Liverpool L69 3BX, UK}
\begin{document}

\begin{abstract}
The coupling $g_{B^\ast B \pi}$ is related to the form factor at zero
momentum of the axial current between $B^\ast$ and $B$
states. Moreover it is related to the effective coupling between heavy
mesons and pions that appear the heavy meson chiral Lagrangian. This
coupling has been evaluated on the lattice using static heavy quarks
and light quark propagators determined by a stochastic inversion of
the fermionic bilinear. We found the value $g=0.42(4)(8)$. Beside its
theoretical interest, this quantity has phenomenological implications
in $B \rightarrow \pi + \bar l l$ decays.
\end{abstract}

\maketitle

\section{INTRODUCTION}

A matrix element with important phenomenological applications is
\begin{equation}
\left\langle B^0(p)\pi ^{+}(q)\mid B^{*+}(p^{\prime })\right\rangle
\label{melem}
\end{equation}
which can be parametrised in terms of the form factor $g_{BB^{*}\pi }(q^2)$ in
the following way:
\begin{equation}
-g_{B^{*}B\pi }(q^2)q_\mu \eta ^\mu (p^{\prime })(2\pi )^4\delta
^4(p^{\prime }-p^{\prime }-q)  \label{gbbp}
\end{equation}
where $\eta ^\mu (p^{\prime })$ is the polarization vector of the asymptotic
state $B^{*+}(p^{\prime })$.

For on-shell external states, $p^{\prime }=p+q$, one can perform an LSZ
reduction of the pion in eq.~(\ref{melem}) 
\begin{equation}
i(m_\pi ^2-q^2)\int e^{iqx}\left\langle B(p)\right| \pi (x)\left|
B^{*}(p')\right\rangle {\mathrm d}^3x  \label{lszred}
\end{equation}
Using the PCAC definition of the pion in terms of the axial current 
\begin{equation}
\pi (x)=\frac 1{m_\pi ^2f_\pi }\partial ^\mu A_\mu (x)  \label{pcac}
\end{equation}
one obtains 
\begin{equation}
q^\mu \frac{m_\pi ^2-q^2}{m_\pi ^2f_\pi }\int e^{iqx}\left\langle
B^0(p)\right| A_\mu (x)\left| B^{*}(p')\right\rangle {\mathrm d}^3x
\label{lszred2}
\end{equation}
Since in the limit $q\rightarrow 0$, eq.~(\ref{lszred2}) and eq.~(\ref{gbbp})
parametrise the same matrix element of eq.~(\ref{melem}), one derives a
Goldberger-Treiman like relation for the pion system~\cite{me} 
\begin{equation}
g_{B^{*}B\pi }(0)=\frac{2m_B}{f_\pi }g  \label{goldberger}
\end{equation}
where $g$ is defined as 
\begin{equation}
g=\int \frac{\left\langle B(p)\right| A_\mu (x)\left|
B^{*}(p)\right\rangle }{2m_B}\eta ^\mu {\mathrm d}^3x  \label{gdef}
\end{equation}

The parameter $g$ is also the effective coupling which appears in the the Heavy Meson Chiral Langrangian~\cite{casalbuoni} and
can be used to constrain the form factors that appear in the matrix elements of the weak vector current, $\overline{u}\gamma ^\mu b$. Which matrix elements are relevant for
$B \rightarrow \pi + \bar l l$ decays. 
One of the best present techniques to extract $V_{{\mathrm ub}}$
is by comparing the experimental values for these form factors with the theoretical ones,
obtained by fitting lattice Montecarlo results. Therefore a precise
determination of $g$ can be used to reduce the number of free parameters in the fit and, therefore, reduce the theoretical uncertainty on $V_{{\mathrm ub}}$.

Even if the matrix element of eq.~(\ref{melem}) does not directly correspond
to a physical process, because $B^{*}\rightarrow B+\pi $ is
kinematically forbidden, the equivalent process for $D$ systems, $%
D^{*}\rightarrow D+\pi $, is indeed physical and the effective coupling 
for this $D$ decay process coincides with our $g$ up to $1/m_c$ corrections.

\section{SIMULATION}

We have performed our simulations on 20 quenched gauge configurations,
generated on a $12^3\times 24$ lattice at $\beta =5.7$, corresponding to $%
a^{-1}=1.10$ GeV, with a tadpole improved SW action ($c_{SW}=1.57$). The
heavy quark propagators are evaluated in the static limit, while the light
quark propagators are evaluated performing a stochastic inversion on the
fermion matrix $Q$%
\begin{equation}
(Q^{-1})_{ij}=\int [{\mathrm d}\phi ](Q_{jk}\phi _k)^{*}\phi _i e^{
-\phi _l^{*}(Q^{\dagger }Q)_{lm}\phi _m}
\end{equation}
and $10$ pseudofermionic fields $\phi _i$ for each gauge configuration were
generated by Montecarlo. Moreover the maximal variance reduction technique
is implemented to reduce the statistical noise~\cite{michael}. Two values of $%
\kappa $ are considered, $\kappa _1=0.13843$ and $\kappa_2=0.14077,$
corresponding to a light quark mass of $140$ MeV and $75$ MeV respectively,
with a critical value $\kappa _{crit}=0.14351$. The interpolating operator 
$J^{\dagger }$ ($J$) that create (annihilate) a static $B$ meson is smeared
with a two-steps fuzzing procedure for the light quark fields. 
We indicate the local operators with the superscript L
and the fuzzed ones with the superscript F.

We have evaluated the two point correlation function at zero momentum 
\begin{equation}
C_2^{FF(FL)}(t)=\overline{\left\langle 0\right| J({\bf y},0)J^{\dagger }(%
{\bf y},t)\left| 0\right\rangle }
\end{equation}
both for two-fuzzed (FF) and one-fuzzed one-local (FL) $J$ operators. The
average is on the spatial position ${\bf y}$ and it has the effect of
increasing the statistics. This has been possible thanks to all-to-all
stochastic propagators. From fitting the asymptotic (large $t$) behavour of $%
C_2$ with a single exponential we have extracted $Z^{L(F)}=\left\langle
0\right| J^{L(F)}\left| B\right\rangle /\sqrt{2m_B}$.

Analogously we have computed the three point correlation function at zero
momentum 
\begin{eqnarray}
C_{3\mu }^{FF}&& \hskip -7mm (r,t_1,t_2)= \nonumber \\
&& \hskip -7mm \overline{\left\langle 0\right| J({\bf y}%
,-t_1)A_\mu ({\bf x}+{\bf y},0)J^{\dagger }({\bf y},t_2)\left|
0\right\rangle }
\end{eqnarray}
for two-fuzzed $J$ operators. Here the average is on both the spatial
coordinates ${\bf y}$ and ${\bf x}$, but keeping fixed $r=\left| {\bf x}%
\right| $. For the spatial indices $\mu =1,2,3$
we also used the rotational invariance of the lattice so that $%
C_{31}=C_{32}=C_{33}$.

Finally we have computed 
\begin{equation}
E^0(r,t)=(Z^F)^2C_{30}^{FF}(r,t,t)/\left[ C_2^{FF}(t)\right] ^2
\end{equation}

and 
\begin{equation}
\overline{E}(r,t)=(Z^F)^2C_{31}^{FF}(r,t,t)/\left[ C_2^{FF}(t)\right] ^2
\end{equation}

\begin{figure}[ht]
\input{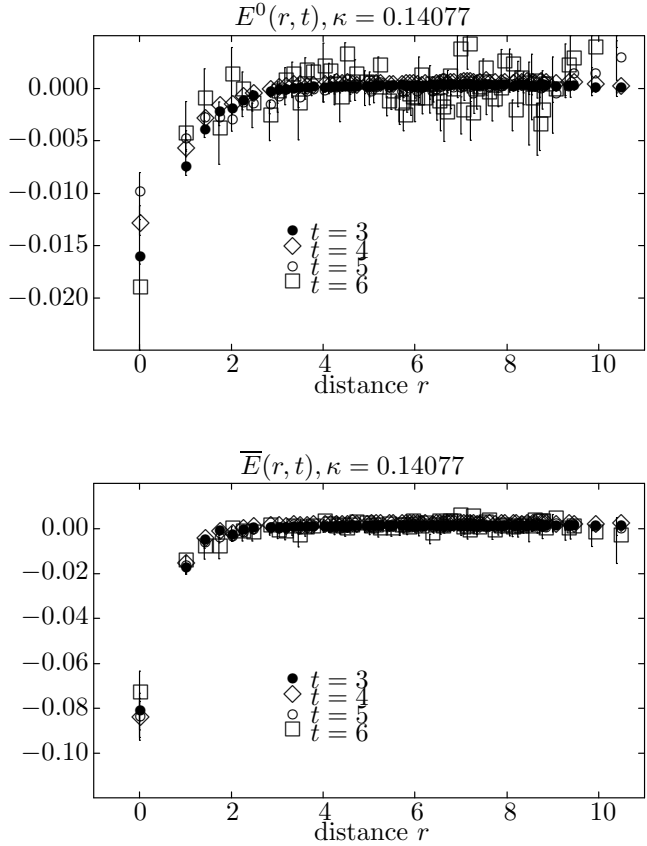}
\caption{Plots of $E^0(r,t)$ and $\bar E(r,t)$ as funtions of $r$ for different values of $t$ ($=3,4,5,6$) for $\kappa_2$.
\label{figure1}}
\end{figure}

These two functions of $r$ are plotted in fig.~\ref{figure1} for
the lightest value of $\kappa $ and values of $t=3,4,5,6$. As expected $%
E^0(r,t)$ $\simeq 0$, because the polarization of a static $B^{*}$ meson is
ortogonal to the temporal direction, while the averaged spatial component $%
\overline{E}(r,t)$ is almost independent on $t$ and presents an exponential
decay in $r$ (this is confirmed by a number of different fitting tests and
by visualizing the points in log scale). In fact in the asymptotic regime in 
$t$%
\begin{equation}
\overline{E}(r,t)=\left\langle B\right| A_\mu (r)\left| B^{*}\right\rangle /%
(2m_B)
\end{equation}

Going back to the definition, eq.~(\ref{gdef}), the lattice regularised $g$
coupling can be extracted from the spatial integral of $f(r)=Se^{-r/r_0}$,
the function fitting $\overline{E}(r,t)$. We find 
\begin{equation}
g^{{\mathrm latt}}=\int f(r)(4\pi r^2){\mathrm d}r=\left\{ 
\begin{array}{ll}
0.54(5) & {\mathrm for }~\kappa _1 \\ 
0.53(5) & {\mathrm for }~\kappa _2
\end{array}
\right.   \label{glatt}
\end{equation}
We observe that the main source of error in $g^{{\mathrm latt}}$ is the
lattice breaking of rotational invariance, in fact the distribution of
the points $\overline{E}(r)$ around the fitting function $f(r)$
exhibits a regular pattern which is independent on $t$ and $\kappa$.

\section{MATCHING}

From $C_2$ we are able to extract the B meson decay constant (in the static approximation), using the relation
\begin{equation}
f_B^{{\mathrm static}}=Z_A^{{\mathrm static}}\sqrt{\frac 2{m_B}}Z^La^{-3/2}
\label{fBstatic}
\end{equation}
For our lattice $Z_A^{{\mathrm static}}=0.78$ is computed using the
Lepage-Mackenzie procedure~\cite{lepage} of matching lattice vs continuum at
a best scale, which in our case is determined to be $q^{*}a=2.29$. Our
result is $f_B^{{\mathrm static}}=0.43(1)(8)$ GeV. We observe that a one-mass
fit gives a value of $f_B^{{\mathrm static}}$ $20\%$ bigger than the value
extracted from a three-masses fit~\cite{michael}. In fact we obtain a value
for this quantity that lies above other lattice calculations of the same
quantity. We take into account this effect by adding a systematic error to
our results (in particular to $g$) of the order $20\%$.

The quantity we are interested in is the $g$ coupling renormalized in the $%
\overline{{\mathrm MS}}$ scheme at the $m_B$ scale 
\begin{equation}
g=Z_A^{{\mathrm tadpole}}g^{{\mathrm latt}}=Z_A^{{\mathrm 1-loop}}\frac{u_0}{u_0^{%
{\mathrm 1-loop}}}g^{{\mathrm latt}}  \label{gmsbar}
\end{equation}
where $Z_A^{{\mathrm 1-loop}}=1-13.8\frac{\alpha C_F}{4\pi }$, $u_0^{{\mathrm %
1-loop}}=1-\pi ^2\frac{\alpha C_F}{4\pi }$ and $u_0=0.86081$ is the average
plaquette. These values imply $Z_A^{{\mathrm tadpole}}=0.806$.

\section{CONCLUSIONS}

Substituting the values of eq.~(\ref{glatt}) into eq.~(\ref{gmsbar}) we obtain 
\begin{equation}
g=\left\{ 
\begin{array}{ll}
0.44(4) & {\mathrm for }~\kappa _1 \\ 
0.43(4) & {\mathrm for }~\kappa _2
\end{array}
\right.
\end{equation}
Performing a naive extrapolation to the chiral limit and including our
evaluation for the systematic error we summarise our result as 
\begin{equation}
g=0.42(4)(8)
\end{equation}
This number has to be compared to the best estimate from a global analysis of
available results~\cite{casalbuoni} 
\begin{equation}
g\simeq 0.38
\end{equation}
and with an estimate obtained from lattice data for the semileptonic B decay
form factor (assuming Vector Meson Dominance), 
\begin{equation}
g=0.50(5)(10)
\end{equation}

A recent, independent, lattice analysis confirmed our result~\cite{pennanen}.

One remark is in order. We consider this study an exploratory one because of
the small lattice, the large lattice spacing, the poor chiral extrapolation
and the use of a non-standard pioneering technique such as stochastic
propagators. As we observed, the $g$ parameter has important
phenomenological applications and we believe our study shows that this
number can be evaluted on the lattice. A better determination is required
and is feasible with present computing power.

\end{document}